\documentclass{ws-p8-50x6-00}

\begin{document}

\title{New Results from NA49}

\author{V.~Friese for the NA49 collaboration}

\address{Fachbereich Physik der Universit\"at Marburg\\ 
Renthof 5, 35032 Marburg, Germany\\
E-mail: volker.friese@physik.uni-marburg.de}

\maketitle

\abstracts{
We present recent results of the SPS experiment NA49 on 
production of strange particles and event-by-event fluctuations
of mean $p_t$ and of charged particle ratios in central Pb+Pb
collisions at various beam energies (40, 80, 158 AGeV) as well as
in different collisions at 158 AGeV, going from p+p over light-ion
collisions to peripheral and central Pb+Pb.}

\section{Introduction}
A possible interpretation of the data collected by the various heavy-ion
experiments at the CERN-SPS over the past few years is that a deconfined
state of nuclear matter is already reached in central Pb+Pb collisions
at top SPS energy (158 AGeV)\cite{qgp}. This immediately triggers
the question whether the transition point can be
experimentally pinned down by varying the collision energy or the size
of the collision system. The experiment NA49\cite{na49} aims to investigate
this question by studying hadronic observables, predominantly strangeness
production, for different beam energies (40, 80 and 158 AGeV) as well as
for different colliding systems, such as p+p, C+C and Pb+Pb at various 
centralities. Further data taking at even lower beam energies (20, 30 AGeV)
is foreseen for the year 2002.

\section{Energy Dependence of Hadronic Observables}

\subsection{Kaon and $\Lambda$ Production}
NA49 can identify charged kaons by time-of-flight measurement near
midrapidity and by the specific energy loss at forward rapidities.
$\Lambda$ baryons are identified in NA49 by their V0 decay topology.
The $m_t$ spectra for both particles types are well described by
exponentials. For the kaons, we find comparable slopes (220-240 MeV)
for all three energies, the slopes for $K^-$ being slightly lower
than those of $K^+$. In the case of $\Lambda$, the slopes increase
slightly with beam energy.

\begin{figure}[t]
 \begin{center}
  \leavevmode
  \epsfxsize=11.8cm
%  \epsfbox{/home/friese/tagung/ismd01/procs/kaon.eps}
  \epsfbox{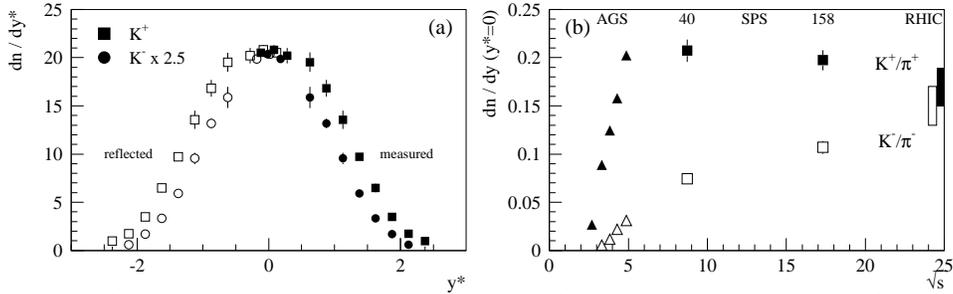}
 \end{center}
 \vspace{-0.5cm}
 \caption{(a) kaon rapidity distributions for
 40 AGeV; (b) energy dependence of the $K/\pi$ ratio at midrapidity. Note
 that the RHIC results are not on horizontal scale.}
 \label{fig:kaon}
\end{figure}

\begin{figure}[b]
 \begin{minipage} [b] {5.9cm}
  \begin{center}
   \leavevmode
   \epsfxsize5.9cm
%   \epsfbox{/home/friese/tagung/ismd01/procs/lambda_rapspec.eps}
   \epsfbox{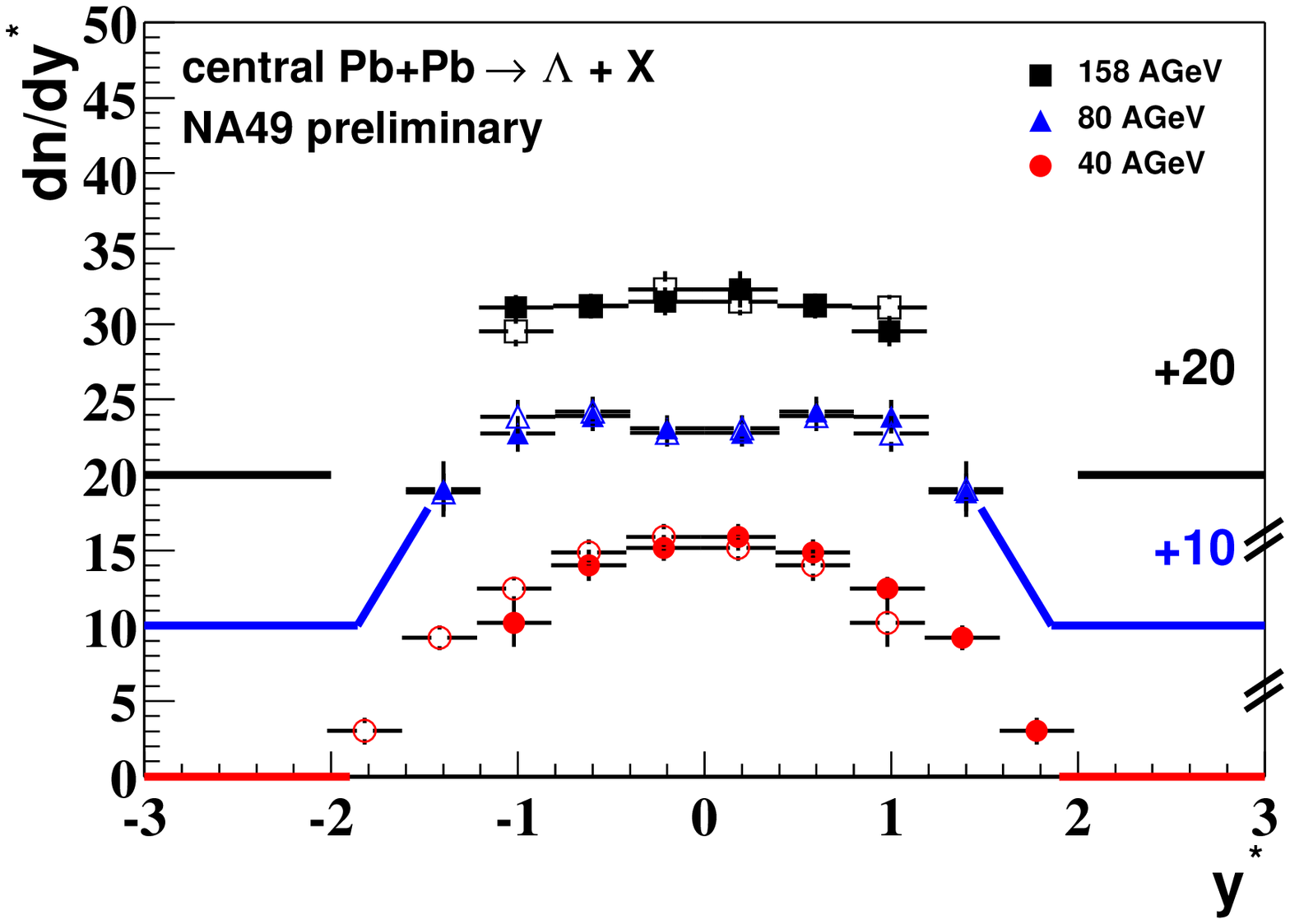}
  \end{center}
 \end{minipage}
 \begin{minipage} [b] {5.9cm}
  \begin{center}
   \leavevmode
   \epsfxsize5.9cm
   \epsfbox{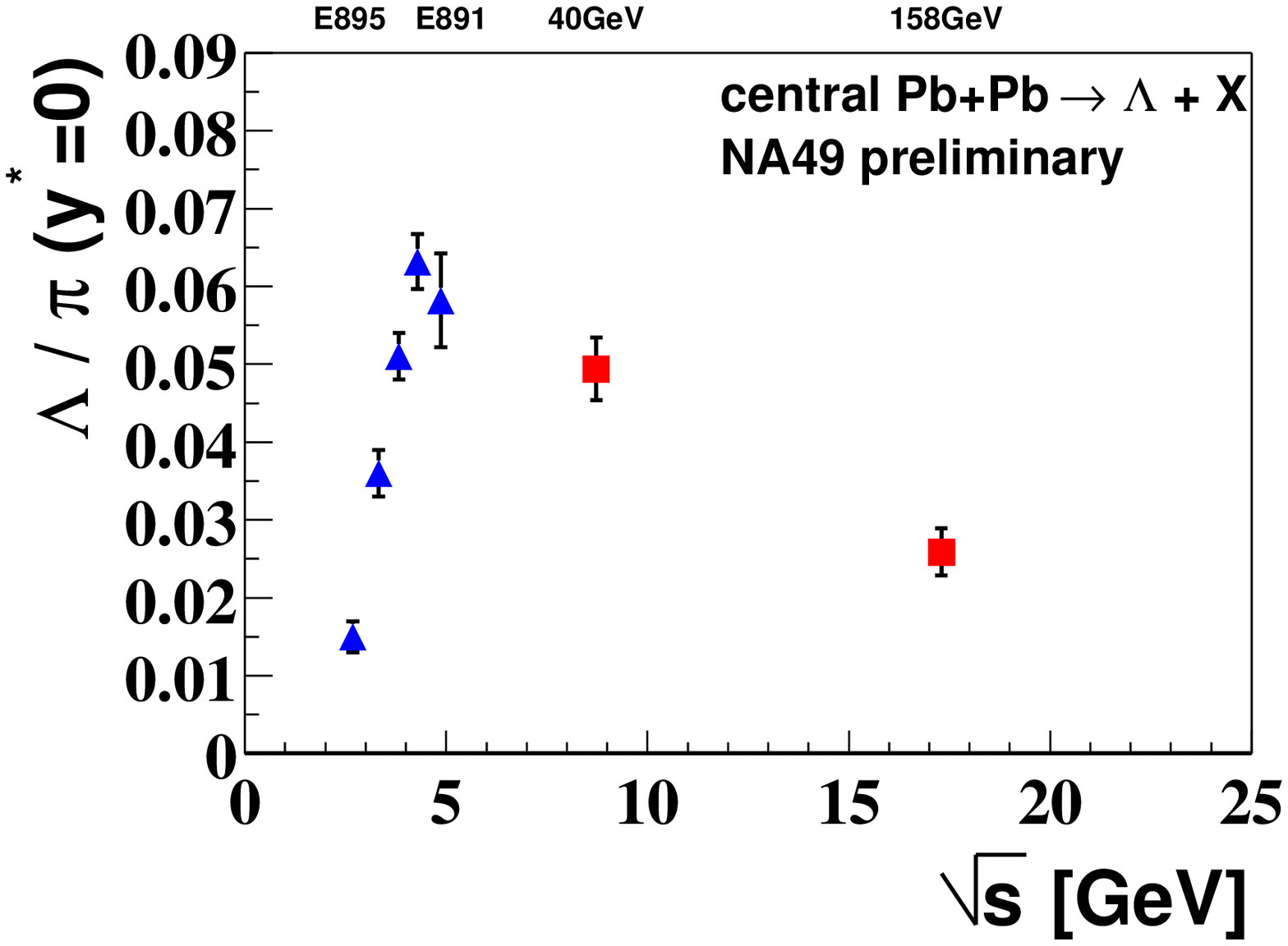}
%   \epsfbox{/home/friese/tagung/ismd01/procs/lam2pi_dndy.eps}
  \end{center}
 \end{minipage}
 \vspace{-1cm}
 \caption{(left) $\Lambda$ rapidity distributions in central Pb+Pb at 40, 80 and
158 AGeV; (right) $\Lambda/\pi$ ratio at midrapidity as function of beam energy. 
AGS data taken
from\protect\cite{agslambda}}
 \label{fig:lambda}
\end{figure}

The rapidity distributions for kaons are shown in fig.~\ref{fig:kaon}a
for 40 AGeV. We observe the $K^+$ to have a somewhat broader distribution
than the $K^-$, which also holds for the other energies. The rapidity
distribution of the $\Lambda$ (fig.~\ref{fig:lambda}) seems to develop a 
plateau around midrapidity when going from 40 AGeV to 158 AGeV. For the latter
case, the shape of the distribution at large $|y|$ is not yet determined,
leaving some uncertainty to the $4\pi$ extrapolation.

Fig.~\ref{fig:kaon}b shows the $K/\pi$ ratio at midrapidity as a function
of $\sqrt{s}$. For 80 AGeV, the analysis of pion production is not yet
finished, so we restrict ourselves to the data at 40 and 158 AGeV.
While a continuous rise in $K^-/\pi^-$ from AGS\cite{agskaon} over SPS to RHIC energies
is observed, the $K^+/\pi^+$ ratio seems to reach a maximum at or slightly 
above top AGS energy. This behaviour is more pronounced in the case of $\Lambda$
(fig.~\ref{fig:lambda}). 

\begin{figure}[bp]
 \begin{minipage} [t] {5.6cm}
  \begin{center}
   \leavevmode
   \epsfxsize5.6cm
   \epsfbox{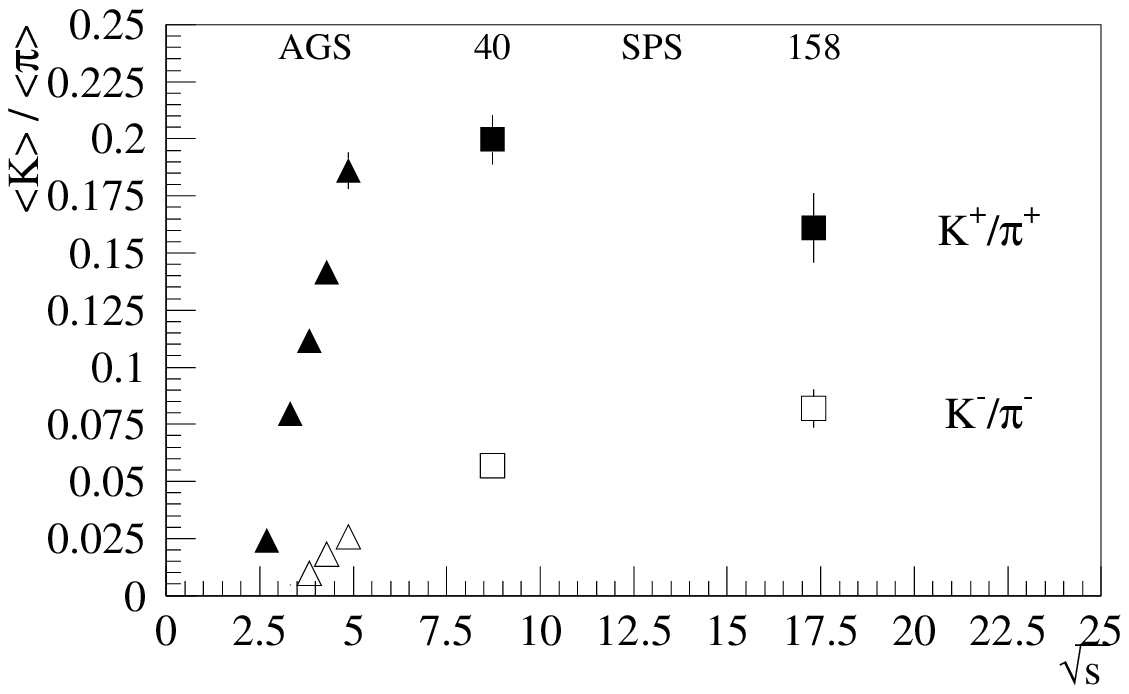}
%   \epsfbox{/home/friese/tagung/ismd01/procs/kaon2.eps}
  \end{center}
  \vspace{-0.5cm}
  \caption{Full phase space $K/\pi$ ratio in Pb+Pb as a function of 
beam energy.}
  \label{fig:kaon2}
 \end{minipage}
 \hfill
 \begin{minipage} [t] {5.6cm}
  \begin{center}
   \leavevmode
   \epsfxsize5.6cm
   \epsfbox{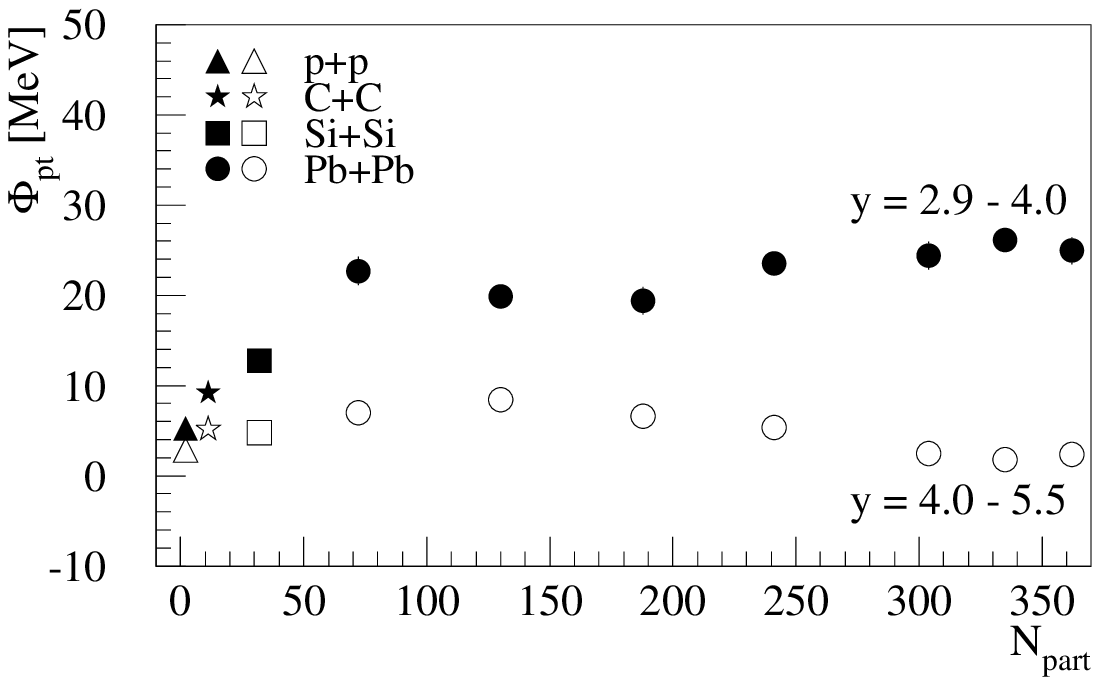}
%   \epsfbox{/home/friese/tagung/ismd01/procs/phipt.eps}
  \end{center}
  \vspace{-0.5cm}
  \caption{$\Phi_{pt}$ as a function of the number of participants at 158
 AGeV for two
  different rapidity ranges}
 \label{fig:phipt}
 \end{minipage}
\end{figure}

The $K/\pi$ ratios in full phase space are shown in fig.~\ref{fig:kaon2}. 
In contrast to $K^-/\pi^-$, the $K^+/\pi^+$ ratio clearly shows
a non-monotonic behaviour.
Most models, including hadron gas models as well as microscopic transport 
models, fail to reproduce such a trend. It is
therefore interesting that the statistical model of the early stage\cite{marek}
indeed predicts a non-monotonic behaviour in the strangeness-to-pion
ratio, assuming a phase transition at about 40 AGeV.

\subsection{Charge Fluctuations}
Event-by-event charge fluctuations have lately been proposed as a
signal of the quark-gluon plasma\cite{charge}. As a measure of these
fluctuations, we have determined $\tilde D = \langle N_{ch} \rangle \langle 
\delta R^2 \rangle / (C_\mu C_y)$, where $\langle\delta R^2\rangle$ denotes
the fluctuations
in the ratio of the numbers of positively and negatively charged hadrons,
as a function of the rapidity window $\Delta y$. 
We obtain similar values around 4 for all
three beam energies, corresponding to the expectation for an
uncorrelated pion gas, whereas a QGP was predicted to yield values 
between 1 and 2.
However, the correction factors $C_\mu$ and $C_y$, which account for the 
finite net charge within the acceptance and global charge 
conservation\cite{koch},
are still under debate; hence at the moment we do not draw a definite
conclusion from the data observed.

\section{System Size Dependence of Hadronic Observables}

\subsection{Kaon and $\phi$ Production}
\begin{figure}[htp]
 \begin{center}
  \leavevmode
  \epsfxsize11.8cm
  \epsfbox{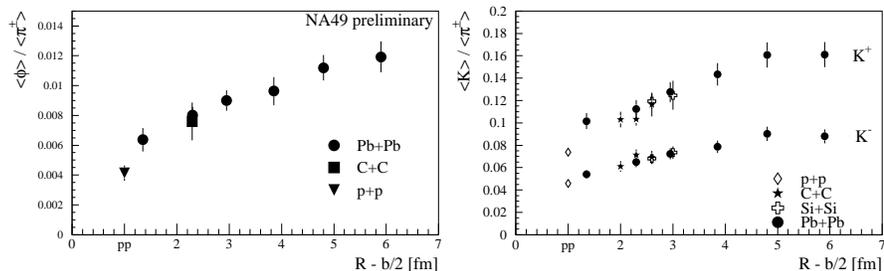}
%  \epsfbox{/home/friese/tagung/ismd01/procs/rb2.eps}
 \end{center}
 \vspace{-0.5cm}
 \caption{$\phi/\pi$ (left) and $K/\pi$ (right) ratios for different collisions types
 at 158 AGeV as a function of $R-b/2$ ($\langle\pi^\pm\rangle
=(\langle\pi^+\rangle + \langle\pi^-\rangle)/2)$}
 \label{fig:rb2}
\end{figure}

The $\phi$ meson is measured in NA49\cite{phipaper} via the invariant mass of its
decay products $K^+K^-$.
When comparing the $\phi/\pi$ ratio in central collisions of light ions
to that in peripheral collisions of heavy nuclei, we find that the
number of participants may not be the right variable to characterise
the reaction, because it does not take into account the collision
geometry\cite{blume}. When using the variable $R-b/2$, with $R$ being the
nuclear radius and $b$ the impact parameter of the collision, we find a smooth evolution in $\phi$ enhancement when
going from p+p over C+C to peripheral and central Pb+Pb. 
The same is true when studying the $K/\pi$ ratio
in p+p, C+C, Si+Si and Pb+Pb at different centralities (fig.~\ref{fig:rb2}).

\subsection{Mean $p_T$ fluctuations}
The observable $\Phi_{pt}$, measuring the non-statistical event-by-event
fluctuations in the mean $p_T$\cite{phipt}, has been studied by NA49
at 158 AGeV for the set of collision types mentioned in the previous
sections. Fig.~\ref{fig:phipt} gives the result as a function of $N_{part}$
for two different rapidity windows. At forward rapidity, we obtain
very small values of $\Phi_{pt}$, almost indepent of $N_{part}$.
Around midrapidity, however, we observe a rise in $\Phi_{pt}$ up to
$N_{part} \approx 100$, from where it stays constant up to central
Pb+Pb. This is contrary to the expectations, which suggest a decrease
in $\Phi_{pt}$ when going from elementary collisions to larger systems
where increasing equlibration suppresses the fluctuations. An 
interpretation of this findings is still lacking.

%\section*{Acknowledgments}
%\input authors.tex

\end{document}